\documentclass[twocolumn,superscriptaddress,aps,prb,amsmath,amssymb]{revtex4} 
\usepackage{bm}
\usepackage{graphicx}

\newcommand{\B}{\mbox{\tiny B}}
\newcommand{\tS}{\mbox{\tiny S}}
\newcommand{\T}{\mbox{\tiny T}}
\newcommand{\SB}{\mbox{\tiny SB}}

\newcommand{\nl}{\nonumber \\}

\newcommand{\be}{\begin{equation}}
\newcommand{\ee}{\end{equation}}
\newcommand{\bsube}{\begin{subequations}}
\newcommand{\esube}{\end{subequations}}
\newcommand{\Eq}[1]{Eq.\,(\ref{#1})}
\newcommand{\Eqs}[1]{Eqs.\,(\ref{#1})}
\newcommand{\Fig}[1]{Fig.\,\ref{#1}}

\newcommand{\la}{\langle}
\newcommand{\ra}{\rangle}

\begin{document}

\title{Hierarchical dynamics for system-bath coherence
      correlation spectrum}

\author{Hou-Dao Zhang}
\author{Jian Xu}
\affiliation{Department of Chemistry, Hong Kong University of Science
and Technology, Hong Kong, China}

\author{Rui-Xue Xu}
\author{Xiao Zheng}
\affiliation{Hefei National Laboratory for Physical Sciences at the
Microscale, University of Science and Technology of China, Hefei
230026, China}

\author{YiJing Yan} \email{yyan@ust.hk}
\affiliation{Department of Chemistry, Hong Kong University of Science
and Technology, Hong Kong, China}

\affiliation{Hefei National Laboratory for Physical Sciences at the
Microscale, University of Science and Technology of China, Hefei
230026, China}

\date{April 29, 2013; to JCP}

\begin{abstract}

  We propose a quasi-particle description for
the hierarchical equations of motion formalism for quantum dissipative
dynamics systems. Not only it provides an alternative mathematical
means to the existing formalism,
the new protocol clarifies also explicitly the physical
meanings of the auxiliary density operators
and their relations to full statistics on solvation bath variables.
Combining with the standard linear response theory,
we construct further the hierarchical dynamics
formalism for correlated spectrum of system--bath coherence.
We evaluate the spectrum matrix for a demonstrative
spin-boson system-bath model.
While the individual diagonal element of the spectrum matrix describes
the system or the solvation bath correlation,
the off-diagonal elements characterize the correlation between
system and bath solvation dynamics.

\end{abstract}

\maketitle

 The hierarchical equations of motion (HEOM) formalism
has been established via the Feynman-Vernon path integral approach
\cite{Tan906676,Xu05041103,Tan06082001,Xu07031107,Din12224103} and
also the stochastic Liouville-equation approach.\cite{Yan04216}
The numerical tractability of this exact quantum dissipation
theory has been extensively demonstrated.\cite{Kre112166,Che11194508,Xu13024106}
It is recognized that the auxiliary density operators (ADOs)
in HEOM contain rich information
on correlated system-bath coherence.\cite{Shi09164518,Zhu115678}
Recently, Shi and coworkers\cite{Zhu12194106}
established an explicit relation between ADOs
and moments of solvation coordinate.
It is also noticed that ADOs
in the conventional HEOM formalism are bosonic in nature,
while those in the second-quantization (electronic)
HEOM are fermionic.\cite{Jin08234703,Zhe121129}  %
This observation implies the possibility
of a quasi-particle picture
that could offer further physical insights on ADOs.

 In this work, we propose the \emph{dissipaton} dynamics
as a quasi-particle approach to
the existing HEOM formalism for bosonic dissipative systems.
Not only it clarifies the physical meanings of ADOs,
the new approach leads also to an
explicit HEOM evaluation for correlated system--bath coherence,
including correlation functions
for solvation bath variables.
Throughout the paper we set $\hbar=1$ and $\beta=1/(k_{B}T)$.
Denote also ${\cal L}(\cdot)\equiv [H_{\tS},(\cdot)]$,
for the reduced system Liouvillian.

 Let us start with the total composite Hamiltonian,
$H_{\T}=H_{\tS}+h_{\B}+H_{\SB}$,
with the system-bath coupling the form of
$H_{\SB} =\sum_{a} Q_{a} F_{a}$.
The system operator $Q_{a}$ here is called a dissipative mode
and can be arbitrary,
while the bath operator $F_{a}$ is called a solvation coordinate,
for its being often modeled as a collection of harmonic bath coordinates.
In the bare bath ($h_{\B}$) interaction picture,
$F^{\B}_{a}(t)\equiv e^{ih_{\B}t}F_{a}e^{-ih_{\B}t}$.
It is a stochastic variable,
characterized by $\la F^{\B}_{a}\ra_{\B}=0$ and
\be\label{C_bath_def}
  C^{\B}_{aa'}(t) \equiv \la F^{\B}_{a}(t)F^{\B}_{a'}(0)\ra_{\B}
= \frac{1}{\pi}\int_{-\infty}^{\infty}\!d\omega
  \frac{e^{-i\omega t}J_{aa'}(\omega)}{1-e^{-\beta\omega}}.
\ee
The second identity is the bosonic fluctuation-dissipation
theorem, with $J_{aa'}(\omega)$ being
the interacting bath spectral density.\cite{Yan05187,Wei08}
The script ``B'' in \Eq{C_bath_def} specifies
the bath ensemble average,
$\la (\cdot) \ra_{\B}\equiv {\rm tr}_{\B}[(\cdot)e^{-\beta h_{\B}}]/
{\rm tr}_{\B}(e^{-\beta h_{\B}})$,
for the dynamical variables in the $h_{\B}$-interaction
picture.
The full-space counterpart, $C_{aa'}(t)$ [cf.\ \Eq{Ct}],
is one of quantities subject to a
direct HEOM evaluation later.

\noindent
\emph{(A) The Dissipaton Approach to HEOM} --
 In line with the HEOM construction,\cite{Tan906676,Xu05041103,Tan06082001}
we decompose $C^{\B}_{aa'}(t)$ of \Eq{C_bath_def},
via certain sum-over-poles scheme, as
\be\label{C_bath0}
  C^{\B}_{aa'}(t) \approx \sum_{k=1}^K \eta^{aa'}_{k} e^{-\gamma^{aa'}_{k} t}.
\ee
It would be exact if $K\rightarrow\infty$.
For clarity we limit our discussion to the real-exponents ($\gamma^{aa'}_{k}>0$)
case such as in an overdamped phonon bath.
The expansion coefficients ($\eta^{aa'}_{k}$) are complex in general.
Thus, each decomposition term
in \Eq{C_bath0} represents
a diffusive mode, which would be classified below as
a \emph{dissipaton}, involved in
the HEOM dynamics of correlated system-bath coherence.

 Introduce the so-called dissipaton operator, $\hat f^{ab}_k(t)$,
having the color-$\gamma^{ab}_k$
and the statistical independence relation
defined for $t>0$ as
\be\label{dissipaton_corr}
   \la \hat f^{ab}_k(t)\hat f^{b'a'}_{k'}(0)\ra_{\B}
  \equiv (\delta_{aa'} \delta_{bb'}\delta_{kk'})
    \eta^{ab}_{k} \exp(-\gamma^{ab}_k t).
\ee
While $\la \hat f^{ab}_k(0^+)\hat f^{ba}_k\ra_{\B}=\eta^{ab}_k$,
the discontinuity at $t=0$  is specified further with
$\la \hat f^{ab}_k(0^-)\hat f^{ba}_k\ra_{\B}=(\eta^{ab}_k)^{\ast}$
and $\la \hat f^{ab}_k \hat f^{ba}_k\ra_{\B}={\rm Re}\,\eta^{ab}_k$.
It is easy to show that individual solvation coordinates,
preserving \Eq{C_bath0}, can now be decomposed as
\be\label{dissipaton_decom0}
  F_{a} = \sum_{b}\sum_{k=1}^K  \hat f^{ab}_k.
\ee
As dissipatons are statistically independent, we can consider them
individually, so that the indices are omitted, i.e.,
$\hat f^{ab}_k=\hat f$ and also for $\eta$ and $\gamma$,
in the following dissipaton approach to HEOM.
We will also exploit the
property of a real-$\gamma$-colored
dissipaton, as defined in \Eq{dissipaton_corr}.
It is diffusive in the pure bath
environment, satisfying\cite{Cha431}  
\be\label{diffusive}
  \text{tr}_{\B}\big[\big(\tfrac{\partial}{\partial t}{\hat f}\big)_{\B}
     \rho_{\T}(t)\big]
 = -\gamma \, \text{tr}_{\B}\!\big[{\hat f}\rho_{\T}(t)\big].
\ee
Here, $\big(\tfrac{\partial}{\partial t}{\hat f}\big)_{\B}=-i[\hat f,h_{\B}]$
and $\rho_{\T}(t)$ is the total system and bath composite density
operator. Its bath trace, $\rho(t)\equiv{\rm tr}_{\B}\rho_{\T}(t) \equiv \rho_0(t)$,
is the reduced system density operator and assigned to
be the zeroth-tier ADO.

 We will show below that $\rho_{n}(t)$,
the $n^{\rm th}$-tier ADO in HEOM,
is related to the $n$-number of \emph{irreducible} dissipatons,
denoted as $\big(\hat f^n\big)^{\circ}$, via
\be\label{rhon_def}
  \rho_{n}(t) \equiv {\rm tr}_{\B} \left[\big(\hat f^n\big)^{\circ}\rho_{\T}(t)\right].
\ee
Introduce also
\be\label{varrho}
  \rho_{n+\underline{m}}(t)
   \equiv {\rm tr}_{\B} \left[\hat f^m\big(\hat f^n\big)^{\circ}\rho_{\T}(t)\right],
\ee
with the underlined $\underline m$ specifying the $m$ dissipatons,
$\hat f^m$, remained reducible.
The Wick's contraction theorem for Gaussian bath leads \Eq{varrho} to
\be\label{Wick}
  \rho_{n+\underline{m}}=\rho_{n+1+\underline{m-1}}
  +n\eta \, \rho_{n-1+\underline{m-1}} \, .
\ee
For the system-bath coupling in the form of $H_{\SB}=Q\hat f$,
\Eqs{diffusive}--(\ref{Wick}), with $\underline{m}=\underline{1}$
and the Liouville-von Neumann equation,
$\dot\rho_{\T}(t)=-i[H_{\tS}+h_{\B}+H_{\SB},\rho_{\T}(t)]$,
lead immediately to
\begin{align}\label{heom0}
 \dot\rho_n &=-(i{\cal L}+n\gamma)\rho_n - i[Q,\rho_{n+1}]
\nl&\quad     - in(\eta Q\rho_{n-1} - \eta^{\ast}\rho_{n-1} Q)\, .
\end{align}
This is just the well-established HEOM,
constructed previously via the
Feynman-Vernon path integral formulations.\cite{Tan906676,Tan06082001,Xu05041103}
The physical meaning of ADOs are also self-evident via
the remarkable relation, \Eq{rhon_def}, to irreducible dissipatons.

\noindent
\emph{(B) White Noise Residue Ansatz} -- It is well known
that a modified HEOM formalism exploits
a white noise residue (WNR)
ansatz.\cite{Tan06082001,Xu07031107,Din12224103}
To obtain the dissipaton prescription of this ansatz
and other related issues hereafter,
it would be sufficient to consider only the case of
$C^{\B}_{aa'}(t)=0$ when $a\neq a'$.
The WNR ansatz starts with the interacting bath correlation function
residue,
\be\label{deltaC_bath}
  \delta C^{\B}_{aa}(t)\equiv
    C^{\B}_{aa}(t) - \sum_{k=1}^K \eta^{a}_{k} e^{-\gamma^{a}_{k} t}
  \simeq 2\Delta_{a}\delta(t).
\ee
Note that $\Delta_a$ is real.
The associated solvation coordinate in dissipatons decomposition
reads now
\be\label{dissipaton_decom}
  F_{a} = \sum_{k=1}^K  \hat f^{a}_k + \delta\hat F_{a},
\ee
with the WNR dissipaton $\delta\hat F_{a}$ being of
\be\label{WNR_dissipaton}
  \la\delta\hat F^{\B}_{a}(t)\delta\hat F^{\B}_{a}(0)\ra_{\B}=2\Delta_{a}\delta(t)
  =\Delta_{a}\lim_{\Gamma\rightarrow\infty}\Gamma e^{-\Gamma t}.
\ee
An important implication here is the {\sc Lemma} that there is
\emph{at most one irreducible white-noise-dissipaton} that can
physically participate in. This {\sc Lemma} will be
verified via self-consistency with its consequences, as seen below.

  Let us introduce the WNR to the case studied in \Eq{heom0},
via setting now $H_{\SB}=Q(\hat f+\delta\hat F)$.
It leads to \Eq{heom0} an additional term,
\be\label{heom1}
   \dot\rho_{n}
 = \big\{\text{terms in \Eq{heom0}}\big\}
  -i[Q,\varrho_{n}],
\ee
with
\be\label{varrho_white_def}
  \varrho_{n}(t) \equiv {\rm tr}_{\B}
    \left[(\delta\hat F)\big(\hat f^n\big)^{\circ}\rho_{\T}(t)\right],
\ee
the one white-noise dissipaton counterpart of \Eq{rhon_def}, satisfying
$\dot\varrho_{n}
 = \big\{\text{\Eq{heom0} with $\varrho$'s}\big\}
- \Gamma\varrho_{n} -i\Gamma\Delta[Q,\rho_{n}]$.
The last two terms
arise from $\big[\tfrac{\partial}{\partial t}(\delta\hat F)\big]_{\B}$
and the contraction of two reducible white-noise dissipatons, respectively.
The contribution from two irreducible white-noise dissipatons
is zero, as inferred from the {\sc Lemma} above.
The convergence of $\dot\varrho_{n}$ in the limit of
$\Gamma\rightarrow\infty$ leads therefore to the relation,
\be\label{final_WNR_1}
  \varrho_{n}(t) = -i\Delta \big[ Q,\rho_{n}(t)\big].
\ee
This is an important result for a white-noise dissipaton.
Substituting \Eq{final_WNR_1} into \Eq{heom1}, we obtain
\begin{align}\label{heom}
 \dot\rho_n &=-(i{\cal L}+n\gamma+\delta{\cal R})\rho_n - i[Q,\rho_{n+1}]
\nl&\quad     - in(\eta Q\rho_{n-1} - \eta^{\ast}\rho_{n-1} Q)\, ,
\end{align}
where $\delta{\cal R}(\cdot) = \Delta [Q,[Q,\,(\cdot)]]$.
We have therefore recovered the modified HEOM formalism,
constructed previously via the Feynman-Vernon path integral formulations.
\cite{Tan06082001,Xu07031107,Din12224103}
The proposed {\sc Lemma} for white noise dissipaton is thus also verified.

 The ADOs in the HEOM formalism
are now completely identified as \Eq{rhon_def}, or
\be\label{rhon_K}
  \rho_{\sf n}(t)\equiv \rho_{n_1\cdots n_K}(t) =
 {\rm tr}_{\B}\!
   \left[\big(\hat f^{n_K}_K\big)^{\circ}\!\cdots\big(\hat f^{n_1}_1\big)^{\circ}
   \rho_{\T}(t)\right].
\ee
For the multiple-dissipative modes case,
each $\hat f^{n_k}_k$ above is understood further as
a collection of $(\hat f^{ab}_k)^{n^{ab}_k}$.
The inclusion of white noise dissipatons does
not add to the ADO indices, but via the
relation of \Eq{final_WNR_1}.
 Moreover, the multiple reducible white noise dissipatons
counterparts of \Eqs{rhon_def} and (\ref{Wick}) read
\begin{align}\label{varrhon_m}
 \varrho_{{\sf n};\underline{m}}(t) &\equiv {\rm tr}_{\B}
    \left[(\delta\hat F)^m\big(\hat f^{n_K}_K\big)^{\circ}\cdots
      \big(\hat f^{n_1}_1\big)^{\circ}
      \rho_{\T}(t)\right]
\nl& =
   \begin{cases}
     \quad \ (\delta\eta)^{j}\, \rho_{\sf n}(t),       &  m=2j
\\
      -i(\delta\eta)^{j} \, \Delta[Q,\rho_{\sf n}(t)],   &  m=2j+1
   \end{cases}\, .
\end{align}
Here, $\delta\eta \equiv \delta C^{\B}_{aa'}(t=0)$, via
the first identity of \Eq{deltaC_bath}.
Apparently, $\varrho_{n;\underline{1}}(t)\equiv \varrho_{n}(t)$
of \Eq{varrho_white_def}.

\noindent
\emph{(C) Statistical Dynamics of Solvation Coordinates} --
Apparently, the zeroth-tier ADO amounts to the reduced system density
operator, $\rho_{\sf n=\sf 0}(t)\equiv \rho(t)= {\rm tr}_{\B}\rho_{\T}(t)$.
The present identification of ADOs as \Eq{rhon_K}
leads to HEOM further for the real-time dynamics
of statistical solvation bath variables.
For example, the moments of solvation coordinates,
in relation to ADOs, can be readily identified,
via \Eqs{dissipaton_decom} and (\ref{rhon_K}),
together with the Wick's contraction theorem of \Eq{Wick}
and its white-noise limit of \Eq{varrhon_m}.

\noindent
\emph{(D) Correlation Function for Solvation Coordinates} --
 Another key result of this work is the establishment
of the HEOM approach to correlation functions for solvation bath variables.
Consider for illustration a two-mode case of
$H_{\SB} = Q_aF_a + Q_bF_b$, in which
$\la F^{\B}_a(t)F^{\B}_b(0)\ra_{\B}=0$, while
$\la F^{\B}_{a'}(t)F^{\B}_{a'}(0)\ra_{\B}
  = \eta_{a'}e^{-\gamma_{a'} t} + 2\Delta_{a'}\delta(t)$.
Therefore,
\be\label{dissipaton_decom2}
  F_{a'}=\hat f_{a'} + \delta\hat F_{a'}; \ \ \  \text{with $a'=a$ or $b$}.
\ee
The corresponding ADOs are therefore
\be\label{rhon_2color}
  \rho_{n_a,n_b}(t) =
 {\rm tr}_{\B}\!
   \left[\big(\hat f^{n_b}_b\big)^{\circ}\big(\hat f^{n_a}_a\big)^{\circ}
   \rho_{\T}(t)\right].
\ee
Turn now to the cross-correlation function for solvation coordinates,
\be\label{Ct}
   C_{ab}(t) = \la F_{a}(t)F_{b}(0)\ra
   = {\rm Tr}\left\{F_a [{\cal G}_{\T}(t)(F_b\rho^{\rm eq}_{\T})]\right\},
\ee
with ${\cal G}_{\T}(t)=e^{-i{\cal L}_{\T}t}$ and
$\rho^{\rm eq}_{\T}=e^{-\beta H_{\T}}/\text{Tr}e^{-\beta H_{\T}}$,
specified in the total system-and-bath composite space.
Let
\be\label{rho_int_b}
   \rho_{\T}(0;F_b)\equiv F_b\rho^{\rm eq}_{\T},
\ee
and $\rho_{\T}(t;F_b)={\cal G}_{\T}(t)\rho_{\T}(0;F_b)$.
Together with \Eq{dissipaton_decom2} for $a'=a$,
we can recast \Eq{Ct} in terms of ADOs as
\be\label{Ct1}
  C_{ab}(t)={\rm tr}_{\tS}\big\{ \rho_{1,0}(t,F_b)
   -i \Delta_a[Q_a,\rho_{0,0}(t;F_b)]\big\}.
\ee
  The initial ADOs for the HEOM evaluation are determined with
\Eq{rho_int_b} for \Eq{rhon_2color}. After
some simple algebra as described earlier, we obtain
\begin{align}\label{rhon_2color_init}
  \rho_{n_a,n_b}(0;F_b) &=
    \rho^{\rm eq}_{n_a,n_b+1} +n_b \eta_b \rho^{\rm eq}_{n_a,n_b-1}
\nl&\quad
  -i\Delta_b[Q_b,\rho^{\rm eq}_{n_a,n_b}] .
\end{align}
The involving thermal equilibrium ADOs are obtained
via the steady-state solutions.
We have thus established the HEOM approach
to correlation functions for solvation coordinates.
As the HEOM evaluation on the system correlation functions
has been well-established,\cite{Zhu115678}
the present development extends its evaluation
for both system and solvation bath dynamical variables.

\noindent
\emph{(E) Numerical demonstrations} -- For demonstration, we consider
a spin-boson system, $H_{\tS}=\frac{1}{2}\epsilon\sigma_z + V\sigma_x$,
with the dissipative mode, $Q=\frac{1}{2}\sigma_z$,
in terms of the Pauli matrixes.
We set $\epsilon=50$\,cm$^{-1}$ and $V=150$\,cm$^{-1}$;
thus the Rabi frequency of the bare system  is
$\Omega_{\rm R}=\sqrt{\epsilon^2+4V^2}=304$\,cm$^{-1}$.
The bath spectral density assumes
$J(\omega)=2\lambda\gamma\omega/(\omega^2+\gamma^2)$,
with $\lambda=150$\,cm$^{-1}$,
but $\gamma=20$, 100, and 200\,cm$^{-1}$,
at $T=77$ and 298\,K, separatively.
We adopt the optimized HEOM formalism\cite{Din11164107}
and the on-the-fly filtering propagator method
that goes with the scaled ADOs.\cite{Shi09084105}
We evaluate $\la F(t)F(0)\ra$
for the solvation coordinate, $\la \sigma_x(t)\sigma_x(0)\ra$
for the spin-system operator,
and the cross-correlations between them, $\la \sigma_x(t)F(0)\ra$
and $\la F(t)\sigma_x(0)\ra$.
Performing the half-Fourier transform on each of them,
\be\label{CABw}
  {\cal C}_{AB}(\omega)\equiv \int_0^{\infty}\!dt\, e^{i\omega t}
  \left[\la A(t)B(0)\ra -\la A\ra\la B\ra\right],
\ee
the spectrum via full-Fourier transform is then
\be\label{SABw}
  S_{AB}(\omega)={\cal C}_{AB}(\omega)+{\cal C}^{\ast}_{BA}(\omega) = S^{\ast}_{BA}(\omega).
\ee
The detailed-balance relation, $S_{BA}(-\omega)=e^{-\beta\omega}S_{AB}(\omega)$,
or its equivalent fluctuation-dissipation theorem,\cite{Yan05187,Wei08}
\be\label{chi_AB}
\chi_{AB}(\omega)\equiv S_{AB}(\omega)-S_{BA}(-\omega)
=(1-e^{-\beta\omega})S_{AB}(\omega),
\ee
is numerically verified in the following converged calculations.
Apparently, $\chi_{AB}(\omega)=\chi^{\ast}_{BA}(\omega)=-\chi_{BA}(-\omega)$.
While $\chi_{AA}(\omega)$ must be real, the off-diagonal $\{\chi_{AB}(\omega)\}$
of interest here are found to be also real, at least numerically.
Consequently, $\chi_{AB}(\omega)=\chi_{BA}(\omega)=-\chi_{AB}(-\omega)$,
for not just the diagonal but also the off-diagonal elements in study.
The converged HEOM evaluation on the aforementioned correlation
functions can therefore be conveniently reported
in terms of the even function $\chi_{AB}(\omega)/\omega$,
with $\chi_{AB}(\omega)/\omega\big|_{\omega=0}=\beta S_{AB}(0)$
[cf.\ \Eq{chi_AB}].

\begin{figure}
\includegraphics[width=0.85\columnwidth]{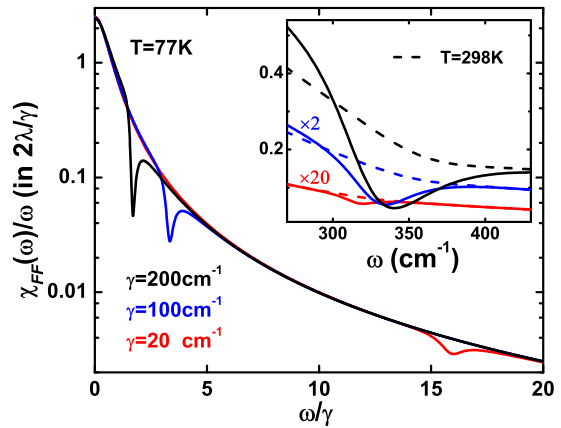}
 \caption{The evaluated $\chi_{FF}(\omega)/\omega$
 at 77\,K, for three values of the bath cutoff frequency,
 $\gamma=200\,{\rm cm}^{-1}$ (black),
 100\,cm$^{-1}$ (blue), and 20\,cm$^{-1}$ (red).
 The inset shows the same function at both 77 K (solid) and 298\,K (dash).
}
\label{fig1}
\end{figure}

\begin{figure*}
\includegraphics[width=0.3\textwidth]{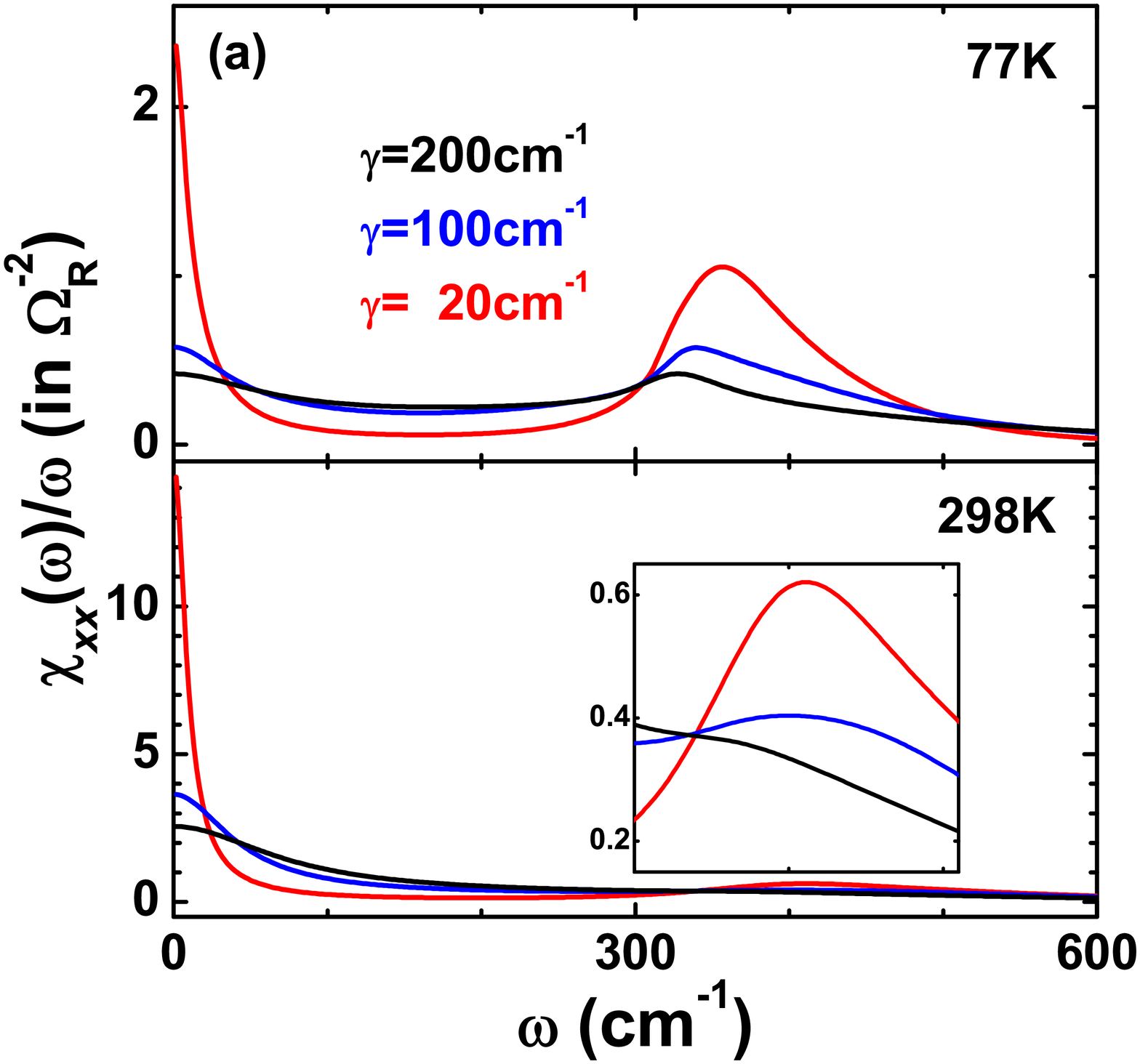}
\includegraphics[width=0.3\textwidth]{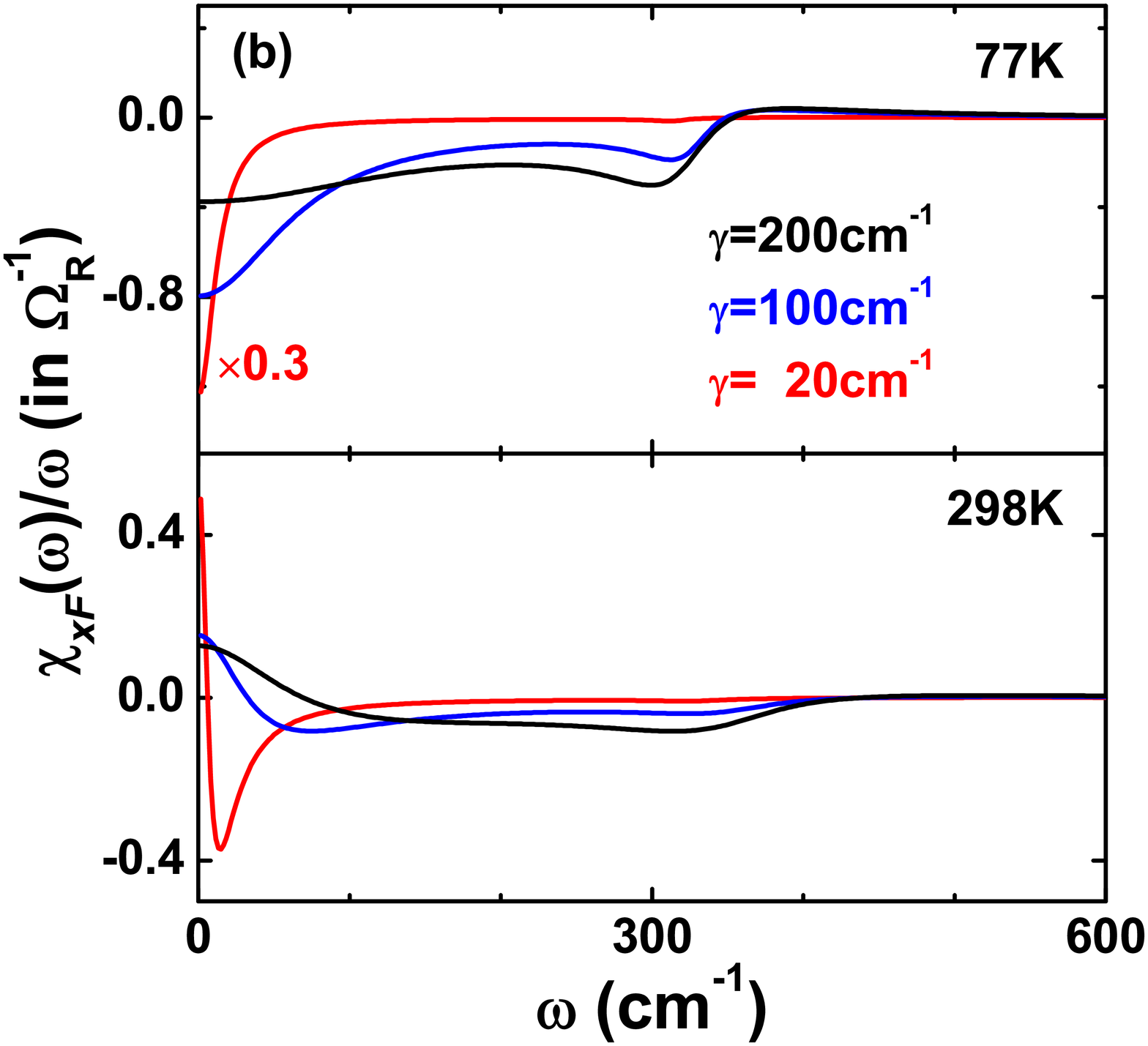}
\includegraphics[width=0.3\textwidth]{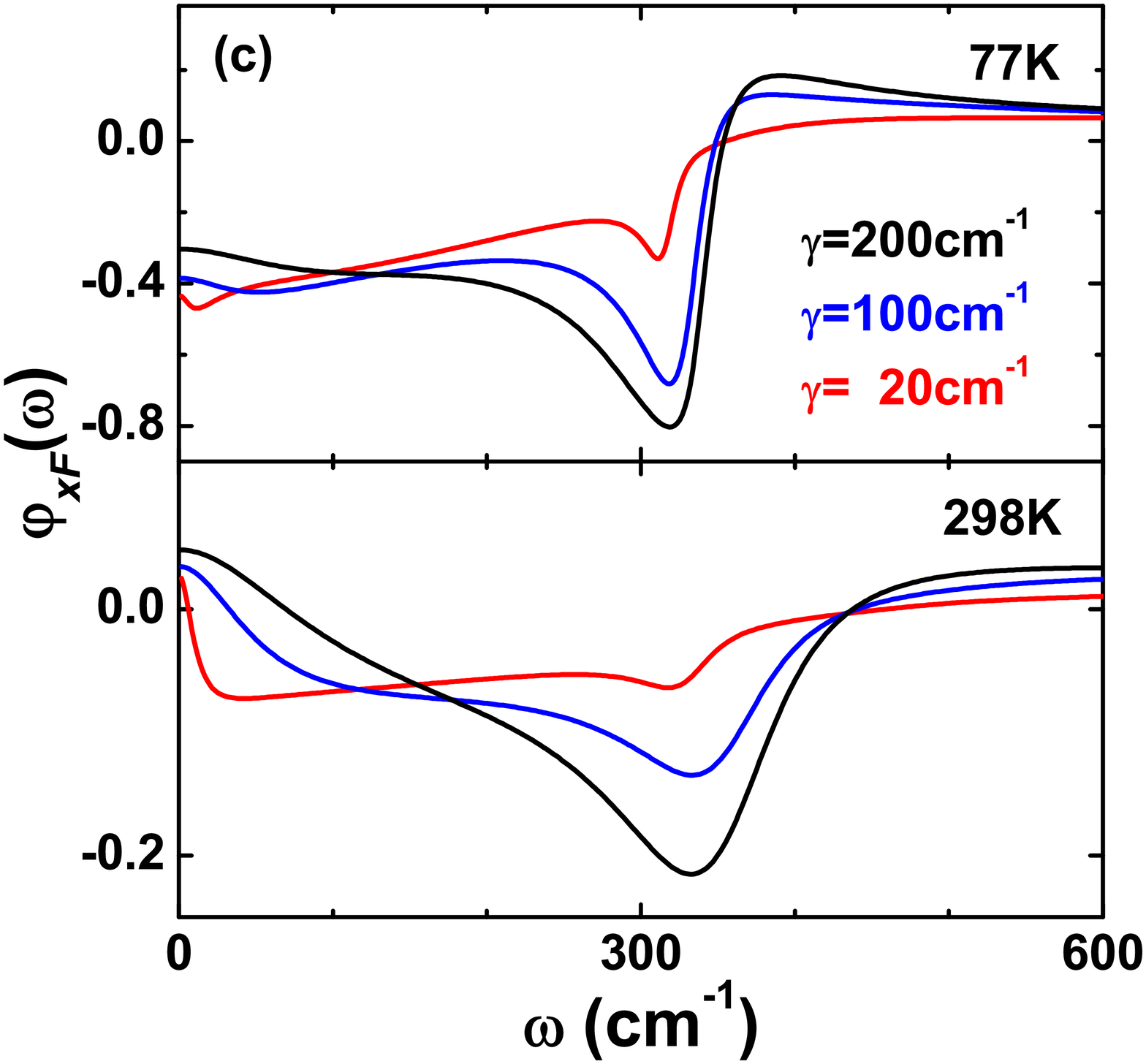}
 \caption{
 (a) The evaluated $\chi_{xx}(\omega)/\omega$ for system operator $\sigma_x$,
 (b) the $\chi_{xF}(\omega)/\omega$ for the cross-correlation
   of system $\sigma_x$ and bath $F$, and
 (c) the coherence spectrum $\varphi_{xF}(\omega)$
     between system and bath,
    at 77\,K (top) and 298\,K (bottom), respectively.
}
\label{fig2}
\end{figure*}

 Figure \ref{fig1} reports the evaluated $\chi_{FF}(\omega)/\omega$
as function of $\omega/\gamma$, at 77\,K.
The bare-bath counterpart of this quantity is
$J(\omega)/\omega = [1+(\omega/\gamma)^2]^{-1}(2\lambda/\gamma)$.
The observed $\chi_{FF}(\omega)/\omega\big|_{\omega\rightarrow 0}>2\lambda/\gamma$
and the dips in \Fig{fig1} indicate
the dominant energy flow from the system to bath in the low frequency regime,
but vice verse in the effective system resonance regime, see also
the inset of \Fig{fig1}.
These features are clearly enhanced as
either $\gamma$ or $\beta$ increases,
along which the system-bath coherence
would increase.

 Figure \ref{fig2}(a) and (b) depict the evaluated
$\chi_{xx}(\omega)/\omega$ and $\chi_{xF}(\omega)/\omega$,
at 77\,K (top) and 298\,K (bottom), respectively,
where $\chi_{xx}\equiv \chi_{\sigma_x\sigma_x}$
and $\chi_{xF}\equiv \chi_{\sigma_xF}=\chi_{F\sigma_x}$.
The observed dependence of $\chi_{AB}(\omega)$
on the $\gamma$ and $T$ parameters
shows a complex interplay
between system-bath coherence and
effective coupling strength in the spectral region
of study. To that end, we present in \Fig{fig2}(c)
the evaluated system-bath coherence spectrum, in terms of
\be\label{varphicoh}
  \varphi_{xF}(\omega)=\frac{\chi_{xF}(\omega)}
  {\sqrt{\chi_{xx}(\omega)\chi_{FF}(\omega)}}\, .
\ee
Note that $|\varphi_{xF}(\omega)|\leq 1$,
for the spectrum positivity.
In the system Rabi frequency region, $\chi_{xx}(\omega)$ would be mainly
controlled by the effective system-bath coupling strength.
A larger $\gamma$ or $T$ would imply a larger effective
system-bath coupling induced dissipation, leading to a smaller
system resonance amplitude, as seen
from \Fig{fig2}(a).
On the other hand, $|\chi_{xF}(\omega)|$ and $|\varphi_{xF}(\omega)|$,
especially  in the Rabi frequency region,
characterize mainly the system-bath coherence,
which increases as $\gamma$ or $\beta$ increases,
as inferred from \Fig{fig2}(b) and (c), and
also the inset of \Fig{fig1}.
The complexity in the vicinity of $\omega=0$
may be understood with the additional complication
arising from the co-occurrence
of peak in the bare-bath $J(\omega)/\omega$ and
that in the bare-system $\chi^{0}_{xx}(\omega)$.
Consequently, the correlated system-bath coherence spectroscopy
shows in general a complex interplay between
the involving system and bath parameters,
the temperature, and the frequency region in consideration.

  In conclusion, the dissipaton picture for ADOs, proposed comprehensively
in this work, clarified the nature
of HEOM for the dynamics in the combined system-solvation bath space.
We identified ADOs be \emph{irreducible} means on
full statistics on solvation coordinates dynamics.
We further addressed issues on the HEOM approach to evaluate
such as the correlation functions for any operators
in the system-solvation bath space.
Thus, the HEOM formalism can be used
directly to extract information
on system-bath entanglement dynamics.
We have also just completed the dissipaton picture  
for fermionic ADOs, together with the HEOM evaluation
on full counting statistics and shot noise spectrum
for transport current, which are experimentally measurable.

   Support from Hong Kong UGC (AoE/P-04/08-2) and RGC (605012),
NNSF of China (21033008 \& 21073169), Strategic Priority Research
Program (B) of CAS (XDB01000000), and National Basic Research Program
of China (2010CB923300 \& 2011CB921400) is gratefully
acknowledged.


\end{document}